\documentclass[usenatbib]{mnras}
\usepackage[T1]{fontenc}
\usepackage{ae,aecompl}

\usepackage{graphicx}	
\usepackage{amsmath}	
\usepackage{amssymb}	
\usepackage{tikz}

\newcommand{\jpb}{{J. Phys. B: At. Mol. Opt. Phys.}}

\newcommand{\rmp}{{Rev. Mod. Phys.}}

\title[Photoionization of Rb II ions]{Single photoionization of the Kr-like Rb II ion in  
                                                        the photon energy range 22 - 46.0 eV}
\author[Brendan M. McLaughlin and James F. Babb]{Brendan M. McLaughlin$^{1,2}$\thanks{E-mail:bmclaughlin899@btinternet.com}
							and James F. Babb$^{2}$\thanks{E-mail:jbabb@cfa.harvard.edu}
\\
$^{1}$Centre for Theoretical Atomic and Molecular Physics (CTAMOP),
           School of Mathematics and Physics,\\
	 Queen's University Belfast, Belfast BT7 1NN, UK\\
$^{2}$Institute for Theoretical Atomic and Molecular Physics (ITAMP)\\
	Harvard Smithsonian Center for Astrophysics, 
	 MS-14, Cambridge, MA 02138, USA\\
		}
\date{Draft: \today}

\pubyear{2017}

\begin{document}

\label{firstpage}
\pagerange{\pageref{firstpage}--\pageref{lastpage}}

\maketitle

\begin{abstract}
{
Single photoionization cross sections for Kr-like Rb$^+$ ions are reported in the energy (wavelength) 
range 22 eV (564 \AA) to 46 eV (270 \AA). Theoretical cross section calculations for this {\it trans}-Fe element 
are compared with measurements from the ASTRID radiation facility in Aarhus, Denmark 
and the dual laser plasma (DLP) technique, at respectively 40 meV and 35 meV FWHM energy resolution. 
In the photon energy region 22 - 32 eV the spectrum is dominated by excitation autoionizing resonance states. 
Above 32 eV the cross section exhibit  classic Fano window resonances features, which are analysed and discussed. 
Large-scale theoretical photoionization cross-section calculations, 
performed using a Dirac Coulomb $R$-matrix approximation are 
bench marked against these high resolution experimental results. 
Comparison of the theoretical work with the experimental 
studies allowed the identification of resonance features and their parameters  
in the spectra in addition to contributions from excited metastable states of the Rb$^+$ ions.
}
\end{abstract}



\section{Introduction}
Rb I optical pumping lines have been 
observed in AGB stars, and both isotopes 
of Rb I ($^{85}$Rb I and $^{87}$Rb I) are 
present based on the presence of s-process elements 
such as Zr \citep{Darling2018}.  
A major source of discrepancy is the quality of the atomic 
data used in the modelling \citep{Sneden1991,Mishenina2002, Roederer2011,Roederer2010,Frebel2014}.

The motivation for the current study of this {\it trans}-Fe element, Rb II,
 is to provide benchmark PI  cross section data for applications in astrophysics.
High-resolution measurements of the photoionization cross section of Rb$^{+}$  
were recently performed at the ALS synchrotron 
radiation facility in Berkeley, California \citep{Macaluso2013}, 
over the photon energy range 22 -- 46 eV at a resolution of 18 meV FWHM.  
Many excited Rydberg states have been identified in the energy (wavelength) 
range 22 eV (564 \AA) to 46 eV (270 \AA). 
Large-scale \textsc{darc} PI cross section calculations when compared  
with previous synchrotron radiation (SR)and  dual laser plasma (DLP) 
experimental studies \citep{Kilbane2007} indicate excellent agreement. 
Such comparisons give confidence in our theoretical data for use in astrophysical applications.

To the authors  knowledge, we are aware only of 
experimental and theoretical work on PI  for this {\it trans}-Fe element, Rb II, 
performed recently at the Advanced Light Source (ALS) \citep{Macaluso2013},
 the joint study by \citet{Kilbane2007}, using a merged-beams technique
 by combining synchrotron radiation (SR) with a beam of Rb$^{+}$ (Rb II) ions 
 at the ASTRID synchrotron radiation facility, the dual laser plasma (DLP) 
technique, at Dublin City University (DCU), and the theoretical study of \citet{Costello2007}. 
This updated earlier work carried out using the DLP technique by \cite{Neogi2003}.

\section{Theory}
\subsection{Atomic Structure}
The \textsc{grasp} code \citep{Dyall1989,Grant2006,Grant2007} generated 
the target wave functions employed in the present work.  All orbitals were 
physical up to $n$=3, 4$s$, 4$p$ and $4d$. We initially used an 
extended averaged level (EAL) calculation for the $n$ = 3 orbitals.  
The EAL calculations were performed on the lowest 
13 fine-structure levels of the residual Rb III ion. 
In our work we retained all the 456 - levels originating from 
one, two and three--electron promotions from the $n$=4 levels into the orbital space of this ion. 
All  456 levels arising from the six configurations 
were included in the \textsc{darc} close-coupling calculation, namely:
$3s^23p^63d^{10}4s^24p^5$, 
$3s^23p^63d^{10}4s4p^6$,   $3s^23p^63d^{10}4s^24p^44d$,
$3s^23p^63d^{10}4s4p^54d$,   $3s^23p^63d^{10}4s^24p^34d^2$, and
$3s^23p^63d^{10}4s^24p^24d^3$.

	%
	%
\begin{table}
\centering
\caption{Rb$^{2+}$ energy levels in Rydbergs (Ry) from the large-scale \textsc{grasp} calculations compared with 
               the available tabulations from the NIST database \citep{NIST2018}.  A sample of the lowest 13 levels 
              for the residual Rb$^{2+}$ ion from the 456-level \textsc{grasp} calculations are shown
              compared to experiment. The percentage difference $\Delta (\%)$ compared to experimentf individual energy 
              levels is given for completeness. \label{tab1}} 
\begin{tabular}{cccccc}
\hline
Level	 &STATE				& TERM		&  NIST	&GRASP		&$\Delta (\%)^a$ \\
	&					&			&  (Ry)	& (Ry)			&			\\		
\hline
1	&$4s^24p^5$			&$\rm ^2P^o_{3/2}$& 0.00000	&0.00000		& 0.0		\\
2	&$4s^24p^5$			&$\rm^2P^o_{1/2}$& 0.06720   &0.06678		& 0.6		\\
\
\\
3	&$4s4p^6$				&$\rm^2S_{1/2}$	& 1.18494  	&1.22287		&3.2		\\
\\
4	&$4s^24p^4({\rm ^3P})4d$	&$\rm^4D_{7/2}$	& 1.40992    &1.44446		&2.5		\\
5	&$4s^24p^4({\rm ^3P})4d$	&$\rm^4D_{5/2}$	& 1.41135    &1.44680		&2.5		\\
6	&$4s^24p^4({\rm ^3P})4d$	&$\rm^4D_{3/2}$	& 1.41761    &1.45368		&2.5		\\
7	&$4s^24p^4({\rm ^3P})4d$	&$\rm^4D_{1/2}$	& 1.42516    &1.46140		&2.5		\\
\\
8	&$4s^24p^4({\rm ^3P})4d$	&$\rm^4F_{9/2}$	& 1.48462  	&1.53650		&3.5		\\
9	&$4s^24p^4({\rm ^3P})4d$	&$\rm^4F_{7/2}$	& 1.50853 	&1.56335		&3.6		\\
10	&$4s^24p^4({\rm ^3P})4d$	&$\rm^4F_{5/2}$	& 1.52929     &1.58192		&3.4		\\
11	&$4s^24p^4({\rm ^3P})4d$	&$\rm^4F_{3/2}$	& 1.53777	&1.59050		&3.4		\\
\\
12	&$4s^24p^4({\rm ^1P})4d$	&$\rm^2P_{1/2}$	& 1.51351	&1.59218		&5.2		\\
13	&$4s^24p^4({\rm ^1P})4d$	&$\rm^2D_{3/2}$	& 1.56776     &1.62083		&3.4		\\
\hline
\end{tabular}
\begin{flushleft}
$^a$Absolute percentage difference, $\Delta (\%)$, between theoretical and experimental energy levels for the Rb III ion.  
The average $\Delta (\%)$ of the energy levels with experiment is $\approx$ 3\%.
\end{flushleft}
\end{table}

Table \ref{tab1} shows the theoretical energy levels from the 456-level 
\textsc{grasp} calculations for the lowest 13 levels of the residual Rb$^{2+}$ ion, compared to the 
values available from the NIST tabulations \citep{NIST2018}.   The average percentage 
difference of our theoretical energy levels compared with the NIST values is 
approximately 3\% higher.  

Photoionization cross sections calculations were carried out on the Rb II ion
for the $3d^{10}4s^24p^6\; {\rm ^1S}_{0}$ ground state, 
$3d^{9}4s^{2}4p^55s\; {\rm ^3P}_{2,1,0}$, $3d^{9}4s^{2}4p^55s\; {\rm ^1P}_{1}$ 
 and the $3d^{9}4s^{2}4p^54d\; {\rm ^3P}_{2,1,0}$  metastable levels 
 using the \textsc{darc} codes. 

	%
	%
\begin{figure*}
\includegraphics[width=\textwidth]{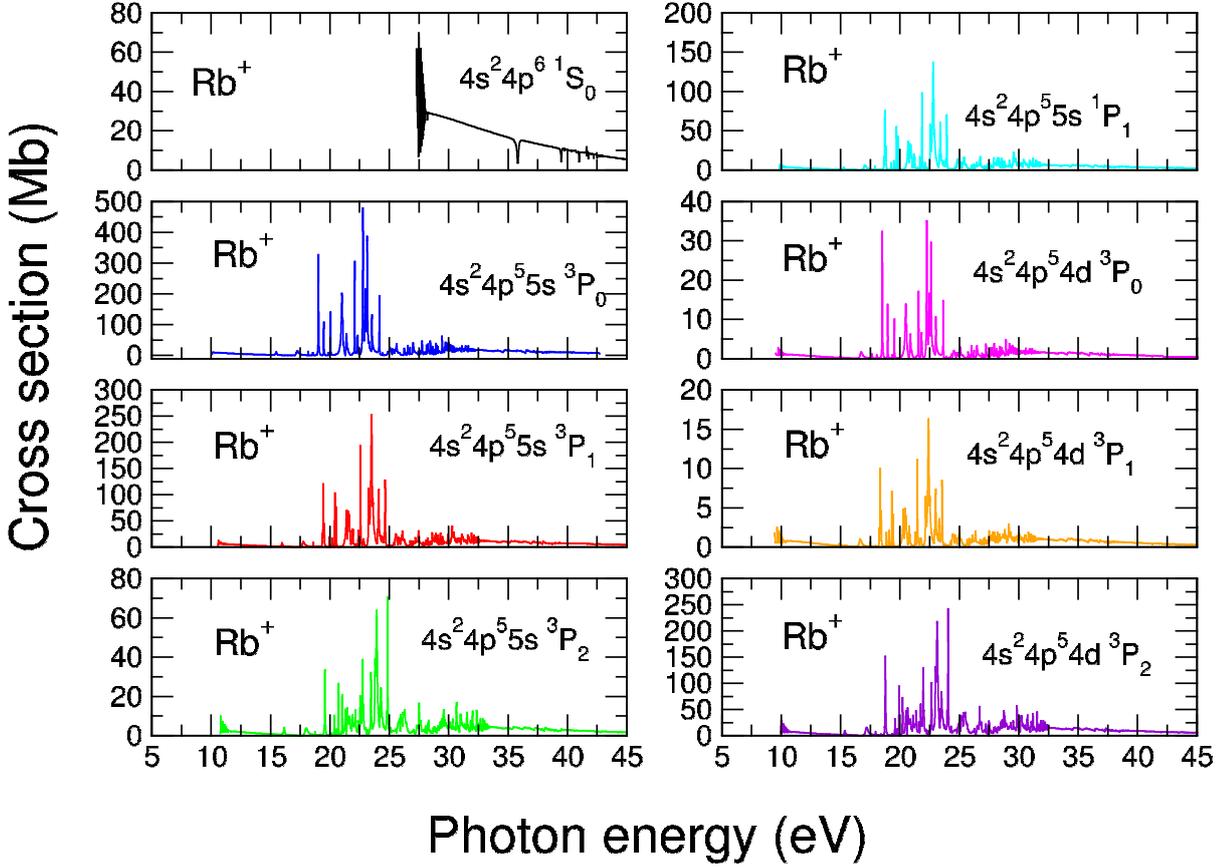}
\caption{(Colour online) Single photoionization cross sections for 
           Rb$^{+}$ ions in the photon energy region 5 -- 45 e . Results
           are illustrated for the 456-level Dirac $R$-matrix approximation, 
           for photon energies from threshold to 45 eV, 
           for the $4s^24p^6\;{\rm ^1S_0}$ ground state,  the 
           $4s^24p^55s\;{\rm ^3P_{2,1,0}}$, 
           $4s^24p^55s\;{\rm ^1P_1}$, and the $4s^24p^54d\;{\rm ^3P_{0,1,2}}$
           metastable levels.  The theoretical cross sections from the 
           456-level \textsc{darc} calculations were convoluted with 
           a Gaussian distribution having a profile width of 18 meV.}
\label{fig1}
\end{figure*}

\subsection{Photoionization calculations}
 The scattering calculations were performed for photoionization cross sections using 
the above large-scale configuration interaction (CI) target wavefunctions  as input
to the parallel \textsc{darc} suite of $R$-matrix codes.
The latest examples of the \textsc{darc} R-matrix method, 
implemented in  our parallel suite of codes 
to predict accurate photoionization cross sections are the 
recent experimental and theoretical studies on the  Zn II trans - Fe ion 
by \cite{Hino2017} and the Ca II ion by \cite{Mueller2017}.

	%
	%
 \begin{table}
  \caption{Resonance energies $E_n$ of the   $4s^24p^{ 5}({\rm ^2P^o}_{1/2}) \ ns\;{\rm ^1P^o_1}$ 
		Rydberg series from the present 456-level \textsc{darc} calculations, 
 		converging to the Rb$^{2+}(3d^{10}4s^24p^5\;\rm ^2P^o_{1/2})$ threshold, originating 
		from the Rb$^{+}(3d^{10}4s^24p^6\;{\rm ^1S_0})$ ground state.
                      The quantum defect $\mu$ for the Rydberg series 
		and linewidths $\Gamma$ ($\mu$eV) are included for completeness.
                      This series is not detectable in the experimental studies due to 
                      the extremely narrow resonance linewidths. \label{tab2}}
\centering
 \begin{tabular}{ccccc}
\hline\hline
  Label				& Theory		&Theory	& Theory	\\
  $ns$   			& $E_n$ (eV)      	&$\mu_n$     & $\Gamma$ ($\mu$eV)\\
\hline\hline
     $8s$			&  --                  	 &  --            	& --             \\
     $9s$   			&   27.4233          	&  0.6502     	&  214        \\
   $10s$   			&   27.5815          	&  0.6493      &  151       \\
   $11s$   			&   27.6960          	&  0.6487     	&  110        \\
   $12s$   			&   27.7816          	&  0.6482   	&  83         \\
   $13s$   			&   27.8472          	&  0.6479  	&  64          \\
   $14s$   			&   27.8986          	&  0.6476   	&  51          \\
   $15s$   			&   27.9397          	&  0.6474     	&  41         \\
   $16s$   			&   27.9730          	&  0.6474    	&  33         \\
\vdots 			&                     	&               	&                \\
  Limit 				&  28.204             	&   --        	& --           \\ 
 \hline \hline                                                               
 \end{tabular} 
 \end{table}

Sixteen continuum orbitals were used in our scattering calculations.
A boundary radius of 9.82 a$_0$ was necessary to accommodate the 
diffuse $n$ = 4  bound state orbitals of the residual Rb III  ion. To 
resolve all the fine resonance features in the spectra, we used an
energy  grid of 2 $\times$ 10$^{-7}$ ${\cal{Z}}^2\;Ry$ (13.6  $\mu$eV), 
where $\cal{Z}$ = 2, in our calculations.  

Photoionization cross section calculations with this 456-level model 
 were carried out using the above energy mesh for the $3d^{10}4s^24p^6\; {\rm ^1S}_{0}$ ground state 
 the  $3d^{9}4s^{2}4p^{5}5s\; {\rm ^3P^o}_{2,1,0}$,  $3d^{9}4s^{2}4p^{5}5s\; {\rm ^1P^o}_{1,}$ 
and  $3d^{9}4s^{2}4p^{5}4d\; {\rm ^3P^o}_{2,1,0}$  metastable levels 
of this ion, over the photon energy range similar to experimental studies.  
	%
	%
\begin{figure*}
\includegraphics[width=\textwidth]{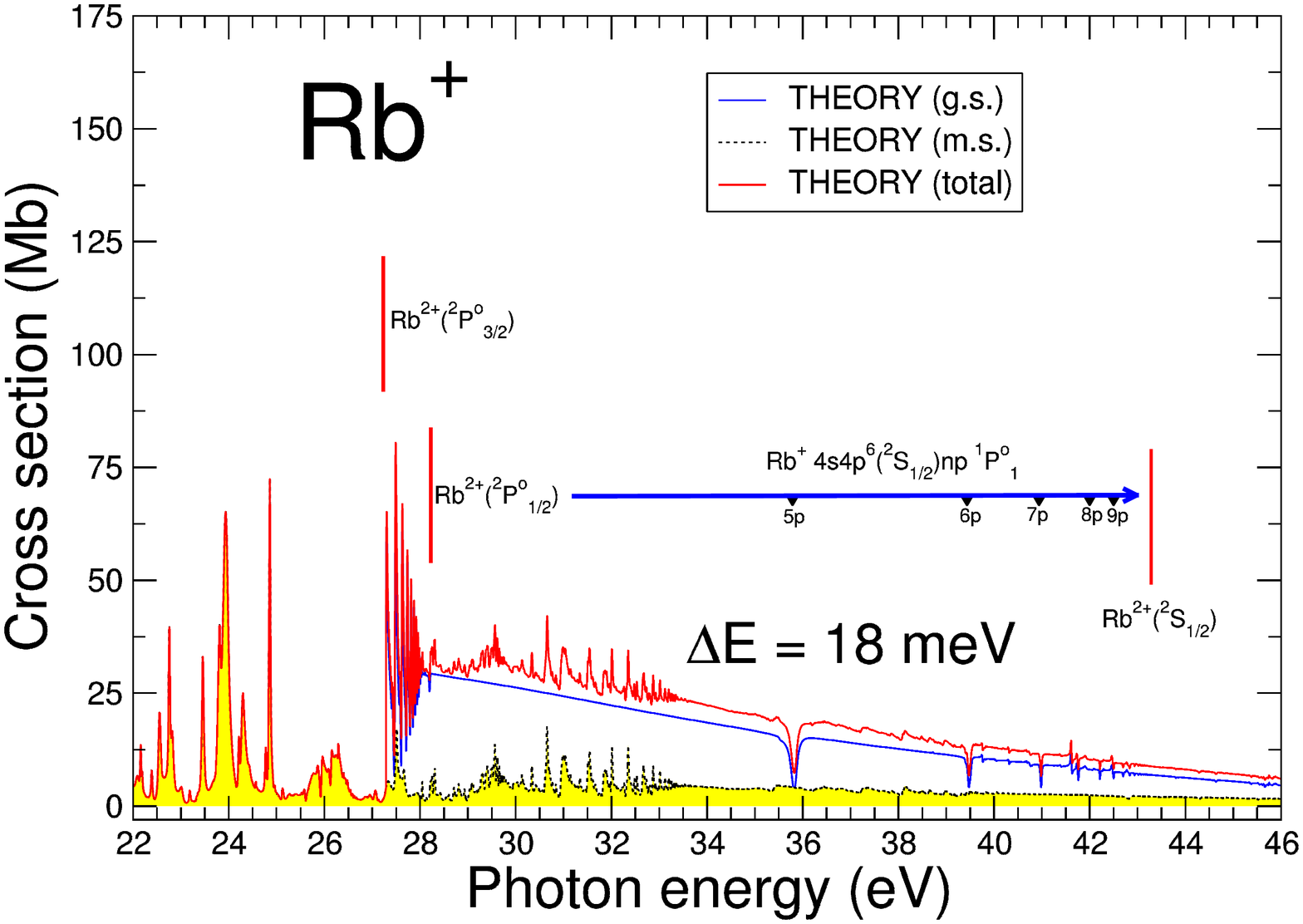}
\caption{(Colour online) Single photoionization cross section of Rb$^{+}$: 
           \textsc{darc} 456-level theoretical results.  
        	The theoretical data were convoluted with 
           a Gaussian distribution having a profile width of 18 meV and an appropriate admixture 
           used (see text for details) for the ground state and the metastable states.
          The Fano window resonances $4s4p^6 ({\rm ^2S_{1/2}})np\; {\rm ^1P^o_1}$ 
           (inverted black triangles) are shown  in the photon energy region 34 eV -- 43 eV.}
\label{fig2}
\end{figure*}

For the ${\rm ^1S}_{0}$ ground state level we require only the bound-free dipole 
 matrices, J$^{\pi}$ = 0$^e$, \ $\rightarrow$ \ J$^{{\prime}\pi^{\prime}}$ = 1$^o$. 
In the case of the  metastable levels we required the following, 
J$^{\rm \pi}$ = 2$^o$, \ $\rightarrow$ \ J$^{{\prime}{\rm \pi}^{\prime}}$ = 1$^e$ ,  2$^e$,  3$^e$,  for 
J$^{\rm \pi}$ = 1$^0$, \ $\rightarrow$ \ J$^{{\prime}{\rm \pi}^{\prime}}$ = 0$^e$ ,  1$^e$,  2$^e$ and
 J$^{\rm \pi}$ = 0$^o$, \ $\rightarrow$ \ J$^{{\prime}{\rm \pi}^{\prime}}$ =1$^e$,  
bound-free dipole matrices.   The Hamiltonian diagonal matrices  were shifted  
to the recommended experimental NIST tabulated  \citep{NIST2018} values. 
Such an energy adjustment  provides better positioning of resonances energies relative 
to all thresholds.
	%
	%
\begin{figure*}
\includegraphics[width=\textwidth]{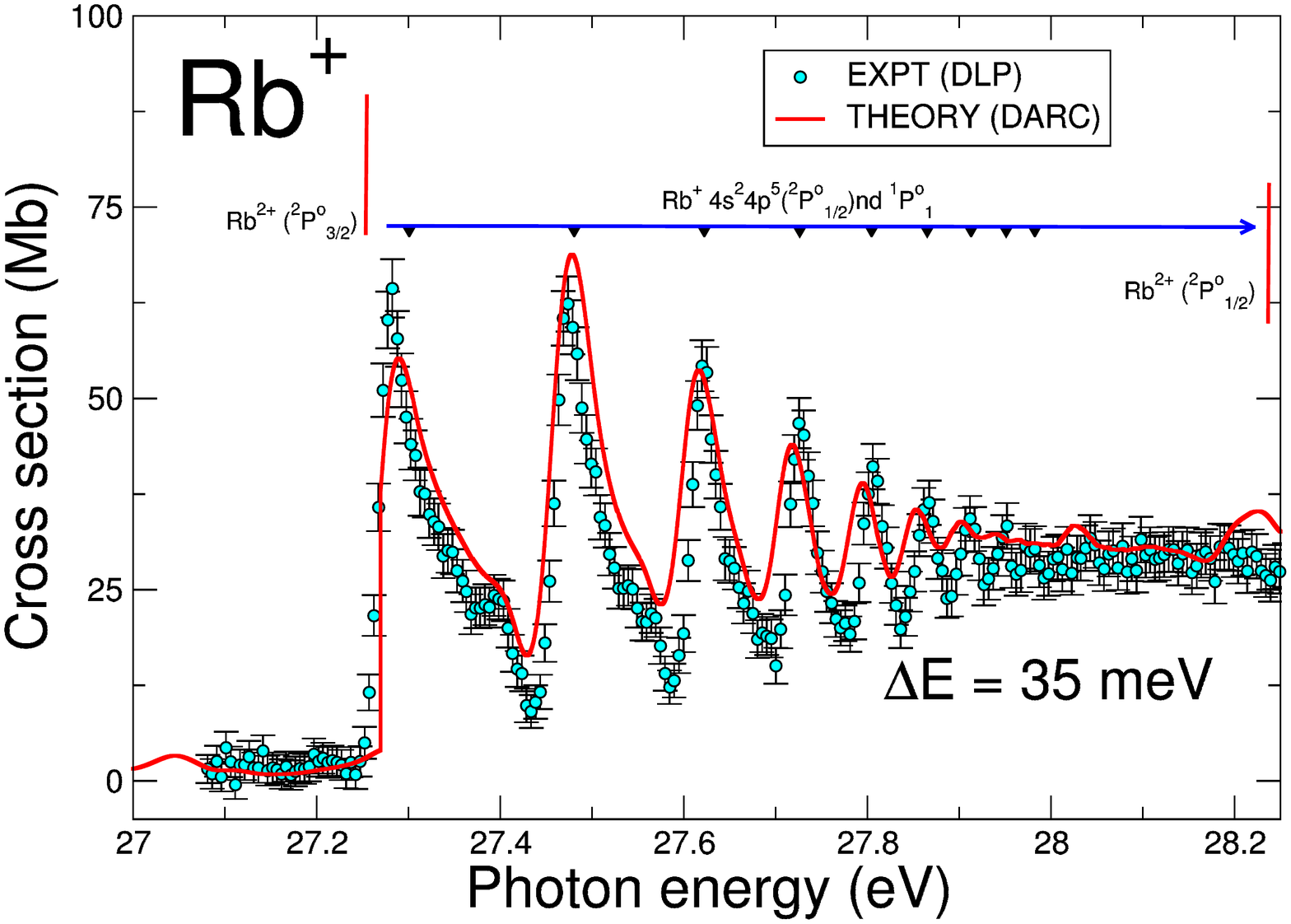}
\caption{(Colour online) 
                 Single photoionization cross section of Rb$^{+}$ in the photon energy region 27 - 28.25 eV.
                 Experimental measurements (solid cyan circles) were obtained using the dual laser plasma technique (DLP) 
                 taken at a photon energy resolution of 35 meV FWHM compared with results
                 obtained from the 456-level \textsc{darc} calculations. The Rb$^+$ $4s^24p^5({\rm ^2P^o_{1/2}})nd\; {\rm ^1P^o_1}$
                 (inverted solid black triangles) are identified in the spectra. Table 3 give a comparison of the \textsc{darc} resonance results 
                  with the DLP experimental  and other theoretical results.
                 The \textsc{darc} photoionization cross sections (solid red line) have been convoluted with a Gaussian
                 distribution having a 35 meV FWHM profile and an appropriate admixture 
                 used (see text for details) for the  ground state and the metastable states.}
\label{fig3}
\end{figure*}

	%
	%
 \begin{table*}
  \caption{Resonance energies $E_n$ of the   $4s^24p^{ 5}({\rm ^2P^o}_{1/2})\ nd\;{\rm ^1P^o_1}$ 
		Rydberg series from the experimental measurements and the 
		theoretical calculations (HXR approximation, Cowan code) of \citet{Kilbane2007} 
		with the present 456-level \textsc{darc} calculations, 
 		converging to the Rb$^{2+}(3d^{10}4s^24p^5\;\rm ^2P^o_{1/2})$ threshold.
                     The Feshbach resonances  in the photoionization cross section originate out of the $4p$-subshell
		from the Rb$^{+}(3d^{10}4s^24p^6\;{\rm ^1S_0})$ ion in its ground state.
                      The quantum defect $\mu$ for the Rydberg series 
		and the autoionization linewidths $\Gamma$ (meV) 
		are included for completeness. \label{tab3}}
\begin{center} 
 \begin{tabular}{ccccccccccccccc}
 \hline \hline  
          \multicolumn{12}{c}{Experimental and theoretical autoionizing Rb II  resonance energies, quantum defects and linewidths}           \\ 
\hline
\\
Label		&Theory$^a$	&Expt$^b$		& Theory$^c$	& Theory$^a$   	& Expt$^b$		&Theory$^c$	& Theory$^a$\\ 
  $nd$   	& $E_n$ (eV)  	 & $E_n$ (eV) 	& $E_n$(eV)    	 &$\mu_n$   		&$\mu_n$  		&$\mu_n$		& $\Gamma$ (meV)\\
\hline
   $7d$   	&     --                   	 &   --         		& --                      	 &    --                  	&  --                      	& --			&   --          \\                  
   $8d$      	&   27.3008              &   27.30  		& 27.3099      	 &  0.2376   		& 0.1977 		& 0.1982		&  32.4      \\
   $9d$      	&   27.4805          	 &   27.50 		& 27.4595     	 &  0.3265      	& 0.2077    		& 0.3922		&  17.2        \\
   $10d$     	&   27.6223          	 &   27.63 		& 27.5900     	 &  0.3257     	& 0.2628		& 0.5853		&  12.3       \\
   $11d$      	&   27.7263          	 &   27.74 		& 27.6814     	 &  0.3253     	& 0.1699 		& 0.7952		&   9.1        \\
   $12d$      	&   27.8046          	 &   27.82 		&27.7519    		 &  0.3246     	& 0.0951		& 1.0283		&   6.9        \\
   $13d$   	&   27.8652           	 &   27.89		& 27.8072  		 &  0.3243   		& -0.1651		& 1.2887		&   5.4        \\
   $14d$   	&   27.9129          	 &   27.95		& 27.8515   		 &  0.3241  		& -0.6377		& 1.5746		&   4.3        \\
   $15d$   	&   27.9512        	 &   --			& --			 &  0.3239     	&  --    		& --			&   3.4        \\
   $16d$   	&   27.9824          	 &   --	 		& --			 &  0.3239    		&  --    		& --			&   2.8        \\
\vdots 	&  	     		 &             		&			 &           		&            		& 			&                \\
  Limit 		&  28.204    	  	 &   28.204    	& 28.204		 &   --   		&  --   			& --			& --     \\ 
 \hline \hline                                                               
 \end{tabular} 
\begin{flushleft}
$^a$Theory, \textsc{darc} 456-level Dirac $R$-matrix approximation.\\
$^b$Experiment, synchrotron radiation (SR) and dual laser plasma (DLP) \citep{Kilbane2007}.\\
$^c$Theory, Hartree-Fock with exchange plus relativistic corrections (HXR) approximation, Cowan code \citep{Kilbane2007}.\\
\end{flushleft}
 \end{center}
 \end{table*}
	%
	%
 \begin{table}
  \caption{Rb$^+ \;4s \rightarrow  np$ Fano window Rydberg resonance parameters for 
                the first few members of the $4s4p^{ 6}({\rm ^2S}_{1/2}) \ np\;{\rm ^1P^o_1}$ series 
                 from the present 456-level \textsc{darc} calculations, and experiment, 
                 converging to the Rb$^{2+}(3d^{10}4s4p^6\;\rm ^2S_{1/2})$ threshold. 
                 Theoretical results obtained from the \textsc{darc} 
                 456-level approximation  are  compared with previous experimental values 
                from the ASTRID synchrotron radiation (SR) facility
                and the dual laser plasma (DLP) technique.
                The Rydberg resonance energy positions $E_r$ are in eV and 
                 the autoionization resonances linewidths $\Gamma$ are in meV. \label{tab4}}
\centering
 \begin{tabular}{cccccc}
\hline\hline
Label			&  Experiment			&  Theory		& Experiment		& Theory \\
  $np$   		& $E_r$ (eV)				&  $E_r$ (eV)   	& $\Gamma$	(meV)   	& $\Gamma$ (meV) \\
\hline\hline
\\
    $5p$		& 35.710 $\pm$ 0.02$^a$	& 35.720$^d$	& 90 $\pm$ 30$^a$	& 137$^d$	\\
     			& 35.708$^b$			& 		     	&  117$^b$          		& 	          \\
			& 35.714$^c$			&			 &  143$^c$			&		\\
\\
    $6p$   		&39.442$^b$            		 & 39.448$^d$    	&  --      	     		& 38$^d$     \\
		 	&39.436$^c$			 &			&			 	&                 \\
\\
   $7p$		    & --					& 40.942$^d$	& --				& 17$^d$ \\
\\ 
 \hline \hline                                                               
 \end{tabular} 
\begin{flushleft}
$^{a}$Dual Laser Plasma (DLP), \citep{Neogi2003}.\\
$^{b}$Dual Laser Plasma (DLP) and Synchrotron Radiation (SR) \citep{Kilbane2007}.\\
$^{c}$Dual Laser Plasma (DLP), \citep{Neogi2003} corrected by \citet{Kilbane2007}.\\
$^{d}$\textsc{darc}, 456-level approximation, present work.\\
\end{flushleft}
 \end{table}
 	%
	%
\begin{figure*}
\includegraphics[width=\textwidth]{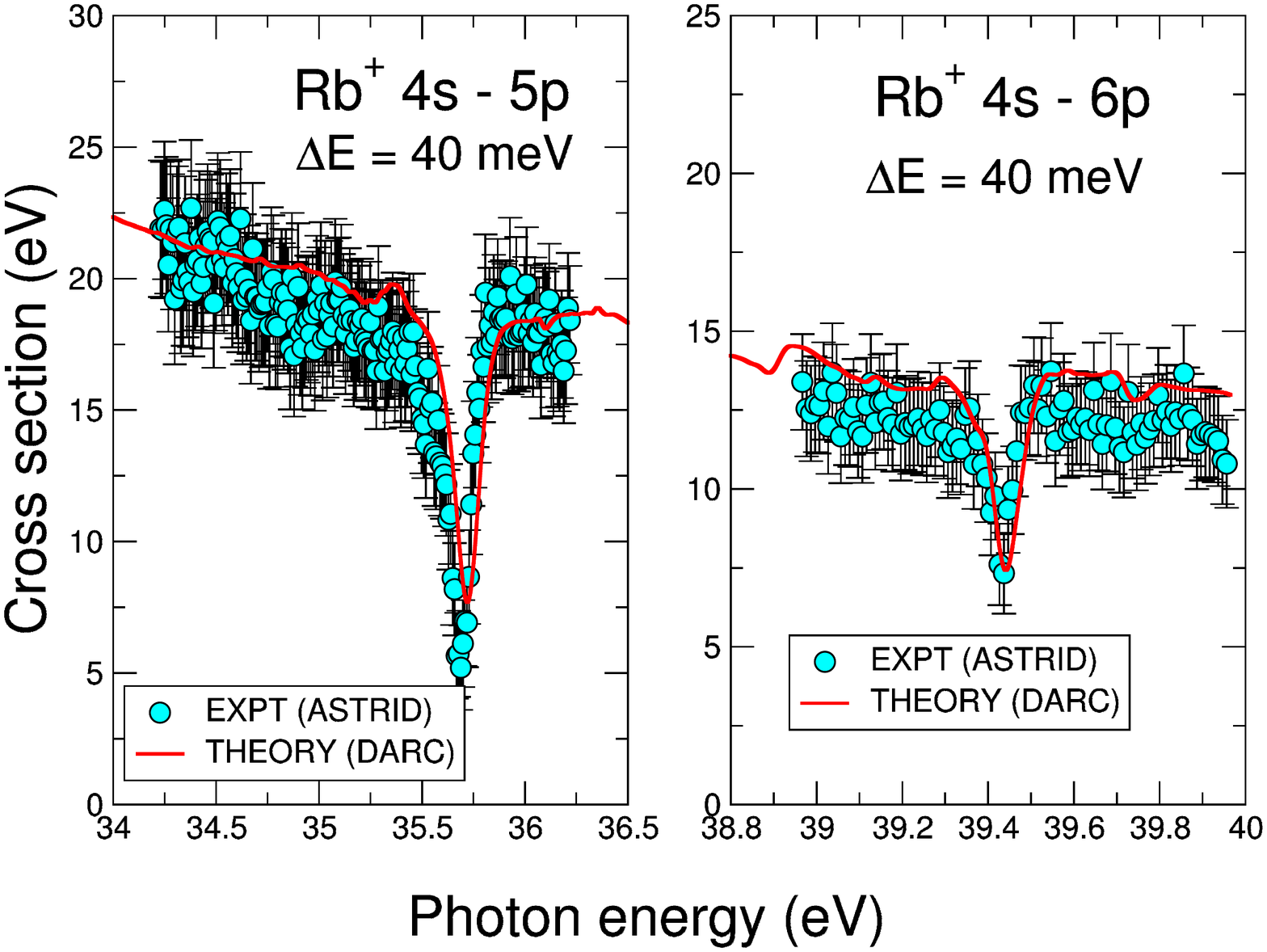}
\caption{(Colour online)  Single photoionization cross section of Rb$^{+}$ 
        in the photon energy regions of the two major Fano window resonances, 
		respectively, 34 -- 36.5 eV and 38.8 -- 40 eV.
        Experimental measurements (solid cyan circles) were obtained from 
        the ASTRID radiation facility, in Aarhus, Denmark,  
        at a photon energy resolution of 40 meV FWHM, co
        mpared with results obtained 
	    from the 456-level (\textsc{darc}) approximation..
        The \textsc{darc} photoionization cross sections (solid red line) have been convoluted with a Gaussian
        distribution having a 40 meV FWHM profile and an appropriate admixture used 
        (see text for details) for the ground and the metastable states.}
\label{fig4}
\end{figure*}

\subsection{Resonances}
The NIST tabulations  \citep{NIST2018} for the Rb II energy levels
were a helpful guidance for the present resonance assignments.
The time-delay of the S-matrix method \citep{Smith1960},
 applicable to atomic and molecular complexes, for locating narrow resonances,
as developed by Berrington and co-workers  \citep{qb1,qb2,qb3} 
was used to locate and determine the resonance positions. This procedure 
was used in our recent study on the trans-Fe element Zn II \citep{Hino2017} to
 locate all the resonances in the spectra.

\section{Results and Discussion}
The 456-level \textsc{darc} PI calculations are shown in Figure \ref{fig1}.
to illustrate the single photoionization cross sections contributions 
from both the ground and the metastable states in
 the region from threshold to 45 eV, 
for the Rb$^+$ ions in their $4s^24p^6\;{\rm ^1S_0}$ ground state, 
the $4s^24p^55s\;{\rm ^3P_{2,1,0}}$, $4s^24p^55s\;{\rm ^1P_1}$ and the 
$4s^24p^54d\;{\rm ^3P_{0,1,2}}$ metastable states. 
We statistically averaged the contribution from the metastable levels and then 
use a best fit with the experimental measurements to identify the metastable 
content in the beam. 

In Figure \ref{fig2} we illustrate our DARC 456-level calculations over the photon energy range 22 -- 46 eV. 
We make the assumption that 98\% of  the  $4s^24p^6\;{\rm ^1S_0}$
ground state and 2\% of the statistically averaged excited metastable states :
 $4s^24p^55s\;{\rm ^3P_{2,1,0}}$, $4s^24p^55s\;{\rm ^1P_1}$ and 
$4s^24p^54d\;{\rm ^3P_{0,1,2}}$  will give the best match with the ALS 
experimental data taken at 18 meV FWHM \citep{Macaluso2013}.

Prior dual laser plasma (DLP) measurements at Dublin City University (DCU) 
made at 35 meV and synchrotron measurements performed on ASTRID 
at 40 meV \citep{Kilbane2007} are compared with our \textsc{darc} calculations. These comparisons 
are  illustrated in figure \ref{fig3} and figure \ref{fig4} respectively. Figure 
\ref{fig3} shows  the comparison of our present \textsc{darc} calculations 
with the DLP measurements taken at 35 meV FWHM in the photon energy 
region 27 - 28.6 where excellent agreement between theory and experiment is seen.  
In figure \ref{fig4}, a comparison with our present \textsc{darc}
PI calculations and the measurements from the ASTRID 
radiation facility in Aarhus (taken at a photon resolution of 40 meV) is made
 in the photon region where the prominent Fano 
window resonances are located.  Here again 
excellent agreement between theory and experiment is observed. 
The good agreement  with the available experimental 
measurements  provides further confidence in our theoretical 
cross section data for astrophysical applications.

From these investigations we see that the PI cross sections below 32 eV are 
dominated by strong Feshbach resonances, while above 32 eV there are  
strong dips in the PI cross section due to Fano window style resonances. 
In the energy range from the ground state threshold to 28 eV, Feshbach 
Rb$^{+}(3d^{10}4s^24p^5ns,md\;{\rm ^1S_0})$ dominate the cross section.
The Fano Rydberg window resonance  series originates from 
photo-excitation of the inner shell  $4s$-electron to the outer $np$ orbital, 
$n \ge 5$,  when the Rb$^{+}(3d^{10}4s^24p^6\;{\rm ^1S_0})$ 
ion is in its ground state. 
The narrow Feshbach  $4s^24p^5ns$ resonances ($\sim$ 0.2 meV or less),
which are tabulated in Table \ref{tab2}, 
are not detectable in the experimental studies 
due to the limited experimental resolution, at best 18 meV, from the ALS work.  
The results for the broader $4s^24p^5nd$ are presented in Table \ref{tab3}, 
where they are compared with previous experimental studies 
\citep{Neogi2003,Kilbane2007}.
Finally, in Table \ref{tab4},  the first few members of the 
Fano window  resonances, $4s \rightarrow np$, $n \ge 5$, 
are tabulated  and their values compared with the 
previous dual laser plasma (DLP) and those from 
the ASTRID synchrotron radiation (SR) facility
experimental studies \citep{Neogi2003,Kilbane2007}. 

The theoretical integrated continuum oscillator 
strength $f$ may also be compared with experiment when available. 
The integrated continuum oscillator strength $f$ of the experimental spectra may be 
calculated over the energy grid [$E_1$, $E_2$], where $E_1$ is 
the minimum experimental energy  and $E_2$ is the maximum experimental energy measured,  
using  \citep{Shore1967,Fano1968,berko1979},
\begin{eqnarray*}
f  &  = & 9.1075 \times 10^{-3} \int_{E_1}^{E_2} \sigma (h\nu) dh\nu  \nonumber \\ 
   & =  & 9.1075 \times 10^{-3} \; \; \overline{\sigma}_{\rm PI}   \\
\end{eqnarray*}
where 
\begin{equation}
 \overline{\sigma}_{\rm PI} = \int_{E_1}^{E_2} \sigma (h\nu) dh\nu
\end{equation}
is the resonance strength.  Evaluating the continuum oscillator strength 
from the  \textsc{darc} theoretical {\it R}-matrix cross sections gave a
value of 3.60 for 98\% of the $4s^24p^6\; {\rm ^1S}_{0}$ ground state
and 2\% of the statistical average of the  $4s^24p^55s\; {\rm ^3P^o}_{2,1,0}$,
 $4s^24p^55s\; {\rm ^1P^o}_{1,}$ 
and $4s^24p^54d\; {\rm ^3P^o}_{2,1,0}$ metastable states.

\section{Conclusions}
Theoretical results from large-scale \textsc{darc} photoionization cross section  calculations were 
used to interpret the experimental data from the DLP and ASTRID facilities.  
From our \textsc{darc} results, a resonance analysis of
both the Feshbach and the Fano window resonances 
illustrate excellent agreement with the available 
experimental data.  The present theoretical work 
may be incorporated into astrophysical modelling codes like 
 CLOUDY  \citep{Ferland1998,Ferland2003}, XSTAR \citep{Kallman2001} 
 and AtomDB  \citep{Foster2012} used to numerically
simulate the thermal and ionization structure of ionized astrophysical nebulae. 
\section*{Acknowledgements}
BMMcL acknowledges support by the US National Science Foundation 
through a grant to \textsc{itamp} at the Harvard-Smithsonian 
Center for Astrophysics under the visitors program, 
the University of Georgia at Athens for the award 
of an adjunct professorship, and Queen's University Belfast for
 the award of a visiting research fellowship (\textsc{vrf}).
\textsc{itamp} is supported in part by NSF Grant No.\ PHY-1607396. 
The hospitality of Professor Thomas J Morgan and  
Wesleyan University, Middletown, CT, USA 
are gratefully acknowledged, where this research was completed.
We thank Captain Thomas J Lavery USN Ret. for his 
constructive comments which enhanced the quality of this manuscript.
Professor John Costello and Dr Deirdre Kilbane, 
are thanked for the provision of the \textsc{astrid} and 
\textsc{dlp} data in numerical format.
The authors acknowledge this research used grants of computing time at the National
Energy Research Scientific Computing Centre (\textsc{nersc}), which is supported
by the Office of Science of the U.S. Department of Energy
(\textsc{doe}) under Contract No. DE-AC02-05CH11231.
The authors gratefully acknowledge the Gauss Centre for 
Supercomputing e.V. (www.gauss-centre.eu) 
for funding this project by providing computing time on the GCS Supercomputer 
\textsc{hazel hen} at H\"{o}chstleistungsrechenzentrum Stuttgart (www.hlrs.de).
\textsc{itamp} is supported in part by NSF Grant No.\ PHY-1607396.
\bibliographystyle{mnras}                       
\bibliography{rbplus}
%
\bsp	
\label{lastpage}
\end{document}